\documentclass[11pt]{article}
\usepackage[utf8]{inputenc}
\usepackage{epsfig}
\usepackage{amsmath,amsfonts,latexsym, amssymb}
\usepackage[T1]{fontenc}

\newtheorem{satz}{Theorem}[section]

\newtheorem{koro}[satz]{Corollary}

\newtheorem{assumption}[satz]{Assumption}

\newtheorem{conclusion}[satz]{Conclusion}
\newtheorem{ob}[satz]{Observation}

\newcommand{\mcal}{\mathcal}

\newcommand{\tit}{\textit}

\begin{document}
\thispagestyle{empty}
\begin{center}
\vspace*{1.0cm}

{\LARGE{\bf An Alternative to Decoherence by Environment and the
    Appearance of a Classical World}} 

\vskip 1.5cm

{\large {\bf Manfred Requardt }} 

\vskip 0.5 cm 

Institut f\"ur Theoretische Physik \\ 
Universit\"at G\"ottingen \\ 
Friedrich-Hund-Platz 1 \\ 
37077 G\"ottingen \quad Germany\\
(E-mail: requardt@theorie.physik.uni-goettingen.de)

\end{center}

\vspace{0.5 cm}

\begin{abstract}
 We provide an alternative approach to the decoherence-by-environment
 paradigm in the field of the quantum measurement process and the
 appearance of a classical world. In contrast to the  decoherence
 approach we argue that the transition from pure states to mixtures
 and the appearance of macro objects (and macroscopic properties) can
 be understood without invoking the measurement-like influence of the
 environment on the pointer-states of the measuring instrument. We
 show that every generic many-body system contains within the class of
 microscopic quantum observables a subalgebra of macro observables,
 the spectrum of which comprises the macroscopic properties of the
 many-body system. Our analysis is based (among other things) on two
 ingenious papers by v.Neumann and v.Kampen. Furthermore we discuss
 the possibility of the emergence of interference among macro states
 via time evolution.

\end{abstract} \newpage
\setcounter{page}{1}
\section{Introduction}
In recent years \tit{environment induced decoherence} has apparently
aquired the status of a new paradigm in the context of the
\tit{quantum measurement process}, the \tit{appearance of a classical
  world}, or the nature of \tit{superselection sectors} (see, just to
mention a few representative sources,
\cite{Zurek1},\cite{Zurek2},\cite{Zeh1},\cite{Schlosshauer}). It is
even sometimes erroneously claimed, that it solves the quantum
measurement problem (cf. \cite{Anderson} and the reply by Adler
\cite{Adler}).

In our view the measurement problem consists of a deeper mystery, that
is, how particular measurement values appear in a single measurement,
and a problem which is of a somewhat lesser calibre, i.e. the
transition of a superposition of states into a corresponding mixture
within the \tit{ensemble picture interpretation} of quantum
mechanics. As papers about this field go into the hundreds, we refer
the reader to the above mentioned reviews what concerns the
decoherence approach, to \cite{Wheeler} as to the older history of the
quantum measurement process and to the recent \cite{Balian} which
discusses the more recent history of the subject. Our main concern in
the following is not to write another review but develop an (in our
view) coherent complementary approach to the above mentioned problems
which is based on a deep (but seldomly cited) paper by v.Neumann, a
later equally important paper by v.Kampen and prior work of the author
(see \cite{Neumann},\cite{Kampen},\cite{Requ2},\cite{Requ3}). See also
the
brief comment of van Kampen in \cite{Kampen2}.\\[0.3cm]
Remark: Only a small part of the content of \cite{Neumann} can be
found in v.Neumann's famous book about the foundations of quantum
mechanics. See \cite{Neumann2} sect. V4 (p. 212ff) where macroscopic
observables are introduced.  \vspace{0.3cm}

One of the reasons why \cite{Neumann} has been largely neglected in
the context of the quantum measurement process is possibly that it is
written in German and that it deals mainly rather with the
ergodic problem. Furthermore, in the fifties (for reasons difficult to
understand) it has been unjustly criticized as being `empty' etc. (a
quite ridiculous remark in our view). As to the reception history see
the recent analysis by Lebowitz et al (\cite{Lebowitz}).

To describe the difference of our appoach compared to the decoherence
appoach in a nutshell one may say: It goes without saying that no
(macroscopic) object is completely isolated, i.e. is in a sense an
open system. This does however not! imply that in an idealized but
perhaps nevertheless reasonable description of nature, we are not allowed
to either neglect these effects or incorporate them in some averaged
statistical manner (as e.g. in the \tit{random phase approximation} in
statistical mechanics, cf. \cite{Huang}). The decoherence by
environment philosophy claims that the entanglement with the
environment is \tit{the} crucial property while we will argue that
e.g. the appearance of \tit{macroscopic objects} can be alternatively
understood in a more intrinsic manner without invoking the (in our
view) rather \tit{contingent} influence of some environment.\\[0.3cm]
Remark: We note that a similar dichotomy exists for example in
statistical mechanics and in particular in ergodic theory. That is,
can ergodicity or statistical behavior only be understood by invoking
some disorder assumption (coming from outside) or can it also be
understood within \tit{closed many-body systems}.\vspace{0.3cm}

There appeared a couple of other investigations which follow a path
similar to ours, i.e. seeking to provide an explanation for the
behavior of quantum measurement instruments which deviates from the
decoherence-by-environment philosophy. Papers we became aware of
belong roughly to two classes. For one there are for example the
papers by Sewell (\cite{Sewell1},\cite{Sewell2}). They
are essentially written in the many-body point of view, described by
us in the beginning of section 3 and which also serves as a basis of
e.g. papers \cite{Requ2} and \cite{Requ3}.               

To the other class belong e.g. papers by Kofler et al
(\cite{Kofler1},\cite{Kofler2}.\cite{Kofler3}). These papers address
slightly different aspects, for example how classicality does emerge
and use as one tool the \tit{Leggett-Garg inequality} (\cite{Garg})
which is about correlations in time instead of space. We will comment
on possible time correlations of macrosystems below in the context of
superposition of macrostates (cf. sect.6). Interesting ideas are also
developed in \cite{Balian}. The view of the authors of \cite{Balian}
concerning the decoherence philosophy is similar to our point of view.

But before we embark on the development of an alternative approach to
the quantum measurement process and the concept of
`\tit{classicality}' in the quantum context, we want to give a very
brief description of the ideas of the decoherence-by-environment
framework as formulated in e.g. \cite{Zurek3}.  
\section{The Decoherence by Environment Concept}
In a nutshell, the idea is quite simple. In the ordinary presentation
of the quantum measurement process (in the v.Neumann spirit) we start
from the following chain of equations. Let $\Phi_0,\Phi_i$ be the (pure) states
of the measuring apparatus, or rather of a subsystem (typically some
macro system). Note however that in the decoherence approach they are
frequently called \tit{pointer states}, the true nature of which is often
not openly specified. Let $\psi_i$ be the eigenstates of the quantum
observable, $A$, to be measured. We then have (with frequently
$\psi_i\Phi_i$ etc. as shorthand for $\psi_i\otimes\Phi_i$ etc.) in a
measurement of $A$:
\begin{equation}\psi_i\Phi_0\rightarrow\psi_i\Phi_i         \end{equation}
The superposition principle of quantum mechanics then yields: 
\begin{equation}\left(\sum_ic_i\psi_i\right)\Phi_0\rightarrow\sum_ic_i\psi_i\Phi_i         \end{equation}

Then follows the argument that the rhs of the last equation cannot be
identified (even in the case of macro objects) with the corresponding
mixture 
\begin{equation}\sum_i|c_i|^2P_{\psi_i\Phi_i}               \end{equation}
$P_{\psi_i\Phi_i}$ being the projector on the state $\psi_i\Phi_i$
irrespectively of the fact that usually
$\left(\psi_i\Phi_i|\psi_j\Phi_j\right)=\delta_{ij}$ is assumed. The reason
  is that in case some of the $|c_i|$ happen to be equal, we observe
  a so-called \tit{basis-ambiguity} (a mathematically coherent
  treatment can be found in \cite{Bub1},\cite{Peres1},\cite{Requ1}).

It is argued in e.g. \cite{Zurek3} and elsewhere that in that cases
there do exist other decompositions of the state vector
$\sum_ic_i\psi_i\Phi_i$ with respect to different bases, which in the
decoherence philosophy can then be associated with different
observables, so that a unique association of macroscopic \tit{pointer
  states} and microscopic states (at first glance) does not seem
possible.\\[0.3cm]
Remark: We give a critical analysis of this point of view in the
following sections.\vspace{0.3cm}

It is said that this stage of the measuring process is only a
\tit{premeasurement} in so far as the state $\sum_ic_i\psi_i\Phi_i$ is
still a pure quantum state being \tit{observably} different from a
mixture! As a typical example the Stern-Gerlach experiment is
frequently invoked where the two split beams can, in principle, be
reunited again into a pure state. This (thought) experiment is
frequently attributed to Wigner (see e.g. \cite{Zurek2}). But it can
already be found in the book by Ludwig (\cite{Ludwig1}) and in an even
earlier interesting paper by Jordan (\cite{Jordan}).  

The measurement process is, according to this philosophy, closed by an
appropriate entanglement of the above state with the so-called
\tit{environment}. If $\varepsilon_i$ are (in the ideal case)
orthogonal states of the \tit{environment}, it is claimed that we
finally have
\begin{equation}\left(\sum_i c_i\psi_i\right)\Phi_0\varepsilon_0\rightarrow\sum_i
  c_i\psi_i\Phi_i\varepsilon_i     \end{equation}
which solves the basis ambiguity (see
\cite{Bub1},\cite{Peres1},\cite{Requ1}).

As a typical example illustrating the basis ambiguity, the following
situation is frequently invoked. The singulett state of two
spin-one-half particles
\begin{equation}1/\sqrt{2}\left((\uparrow)(\downarrow)-(\downarrow)(\uparrow)\right)         \end{equation}
can be represented in e.g. the eigen basis of the $x$-component of the
spin, i.e. 
\begin{equation}\psi_1'=\left((\uparrow)+(\downarrow)\right)/\sqrt{2}\quad
    ,\quad
    \psi_2'=\left((\uparrow)-(\downarrow)\right)/\sqrt{2}        \end{equation}
as
\begin{equation}-1/\sqrt{2}\left(\psi_1'\psi_2'-\psi_2'\psi_1'\right)     \end{equation}
Remark: Note that the exact compensation of the other cross terms come
about because of the common prefactor $1/\sqrt{2}$, i.e. the necessary
and sufficient condition for a basis ambiguity mentioned
above.\vspace{0.3cm}

In the decoherence philosophy according to e.g. Zurek the second
tensor product component may then be associated with some pointer that
is (part of) a measurement instrument. It is then argued that in the
three-orthogonal Schmidt-decomposition , $\sum_i
c_i\psi_i\Phi_i\varepsilon_i$, which is a delocalized state due to the
structure of the environment states, $\varepsilon_i$, one can locally
regard the measurement outcome as a mixture, $\sum_i
|c_i|^2P_{\psi_i\Phi_i}$, (by tracing over the environment) while the
global state is still a pure vector state with the information spread
into the environment. That is, the crucial point is that one remains
globally in the regime of unreduced quantum states!

In our view, at least two of the conceptual ingredients are
problematical. First, in case of e.g. the Stern-Gerlach experiment,
which serves as kind of a paradigm, the (silver) atoms, carrying the
spin degree of freedom, are frequently regarded as pointers
(cf. e.g. \cite{Zurek3}), or at least as a similar device. In our
view, and in the original quantum measurement literature (see
e.g. \cite{Braginsky1}), one would rather call such a subsystem a
\tit{quantum probe} in the context of \tit{quantum non-demolition}
measurements. A similar role is played by the photon in the
\tit{quantum microscope}. In general it may be subsumed under the
catchword of \tit{shift of the cut} between the micro and the macro
world in the measurement process. To put it briefly, we have the
impression that important subsystems like photo plates, magnets, and
the like, which we would prefer to regard as essential parts of the
measurement instrument are now simply called environment in the
decoherence approach. This exactly happens in the paradigmatical
Stern-Gerlach experiment.

In \cite{Zurek3} the measurement interference is for example
idealialized to a \tit{two-bit which path} measurement. We think, the
situation is not satisfactorily analysed in parts of the decoherence
literature. Either the detector is quantum, then the measurement
interference is not! \tit{closed} (almost by definition) and no
decision is made. Or the decision is made irreversible, this can only
happen if the detector is a macroscopic device like e.g. a
photographic plate. In that case it is no longer a premeasurement as
our following analysis shows. In any case, no environment is really
envolved in the process.

Second, it is claimed that the \tit{pointer basis} (and ultimately the
correct functioning as a measurement instrument) is established via
the interaction of the pointer states with the environment. We must
say that we are extremely sceptical, if this point of view is really
correct and we will substantiate our scepticism below. We rather think
that pointer basis and functioning as a measurement instrument are a
priori fixed by the concrete setup of the instrument according to some
pre-theory of measurement, typically incorporating pieces of classical
and quantum physics. This we can at least learn from the analysis of
concrete measurement situations (\cite{Braginsky1}) and the work of
the founding fathers of quantum theory (cf. the beautiful discussion
between Einstein and Heisenberg as described in \cite{Heisenberg2}). A
typical ingredient is usually some sub-system being in a meta-stable
state (photo plate, Wilson chamber, spark chamber etc.).

Third, the ordinary environment is usually of a very contingent
character and it is at least debatable to attribute pure quantum
states to it, and, a fortiori, states which are assumed to play a role
relative to the pointer states as do measuring instruments
relative to the micro objects (in the words of Zurek in
\cite{Zurek2}). We would like to emphasize that the interaction of a
measuring instrument with a micro object is a very special one while
the interaction of a pointer with the environment is usually of the
ordinary statistical type.

On the other hand, the influence of the environment has played an
important role already in the classical literature about the quantum
measurement process (cf. e,g, the lucid analysis of Heisenberg in his
contribution to the Bohr-Festschrift \cite{Heisenberg1}). He clearly
states that an apparatus, not interacting with the exterior world, is
a quantum system and cannot be used as a measuring instrument. It is,
in his words, in a potential, i.e. a quantum state. It becomes a macro
system via its contact with the environment (thus aquiring factual
properties). Furthermore, the illustrations of concrete measuring
instruments in the contribution to the Bohr-Einstein debate in
\cite{Wheeler}, with their solid clamps and bolts clearly show that a
strong contact with the environment is important.

As a last point, the influence of the environment is also incorporated
in statistical mechanics. Starting from  a global pure state
(system plus environment) it is shown in e.g. \cite{Huang} how one
arrives via the \tit{random phase approximation} at a statistical
state of the system. That is, it is not the influence of the
environment which is denied by us but rather the ubiquity of the
invoked \tit{measurement-like} effect on the pointer states and its
role for the appearance of a classical world.
\section{Macro Observables from Quantum Theory}
In this section we descibe in a, as we think, coherent way how
\tit{macro observables} and macroscopic properties do emerge
\tit{within} the framework of quantum theory. The description is based
on the highly original papers by v.Neumann and v.Kampen
(\cite{Neumann},\cite{Kampen}), some related work of Ludwig
(\cite{Ludwig2},\cite{Ludwig3}) and prior work of the author
(\cite{Requ2},\cite{Requ3}).

Most of \cite{Ludwig2},\cite{Ludwig3},\cite{Requ2},\cite{Requ3} is
written in the many-body-language approach to the measurement process
with relations to phase transitions and super selection
sectors. Papers written in a similar spirit are e.g. from the italian
school (see for example \cite{Danieri}) or the papers by Sewell
already mentioned above. A central problem discussed in these papers
was the treatment of macroscopic systems as quantum systems, a problen
which also troubled Legett (see e.g. \cite{Legett1}). We think, a
transplantation of the above ideas of v.Neumann and v.Kampen into this
measurement context will clarify some longstanding open questions. That
is, we will show in the following how the macroscopic regime is
embedded as a subtheory in general quantum physics.

In this section, for short, we will mainly discuss the ideas of v.Kampen. We
start from a many-body wave function
\begin{equation}\Psi(q)=\sum a_n\psi_n(q)\,e^{-iE_nt/\hbar}\quad ,\quad
  q=(q_1,\ldots,q_f)     \end{equation}
with $\psi_n(q)$ the eigenfunctions of the microscopic Hamiltonian,
$H$. It is not really necessary to discuss the distribution of
spectral values of $H$ in any detail. We know that for $f\gg 1$ they
are usually irregularly distributed in dense clusters (at least for
ordinary many-body systems) and are also typically (highly)
degenerated. We postpone the discussion of more
particular systems (displaying e.g. socalled macroscopic quantum
phenomena) to forthcoming work (cf. also the discussion in
\cite{Legett1}). Put differently, there may exist particular many-body
systems (or states) which have a more pronounced quantum nature,
i.e. have  a more regular spectrum, but in this paper we will
concentrate on systems with the usual macroscopic properties. 
The crucial idea is the existence of what v.Neumann and
v.Kampen call \tit{macroscopic observables} (a possible construction
is given in e.g. \cite{Neumann}), other constructions are given in
\cite{Ludwig3} or \cite{Requ2},\cite{Requ3}; we come back to this
point below, in particular in the last section. 

In the following we mainly use the notation of v.Kampen.
\begin{ob}There exist (almost) commuting observables in the representation
 space of the many-body system, denoted by $E,A,B,\ldots$ ($E$
 representing the macroscopic, i.e. coarse-grained energy operator)
 and a complete, orthonormal set of
 (approximate) common eigenvectors, $\Phi_{Ji}$, with the property
 \begin{equation}A\circ\Phi_{Ji}=A_J\cdot \Phi_{Ji}+O(\Delta
   A) \end{equation}
 where $\Delta A$ is the measurement uncertainty
 of the macro observable $A$. It is always assumed that $\Delta A$ is
 macroscopically small but large compared to the quantum mechanical
 uncertainty $\delta A$. The approximate common eigenvectors come in
 groups, indexed by $J$ with $i$ labelling the vectors belonging to
 the group $J$.
\end{ob}
The above equation is assumed to hold for all macro observables. The
subspace, belonging to $J$ is called a \tit{phase cell}. It is assumed
that the eigen values $A_J$ are macroscopically discernible, i.e. they
describe different macroscopic behavior. That is, quantum states 
belonging to the same phase cell have the \tit{same} macroscopic
properties but are microscopically different. This is, by the way, the
same concept as the concept of phases and superselection sectors in
the above mentioned many-body approach to the quantum measurement
process. That is, the phase cells will go over into the latter
concepts after a certain idealization (thermodynamical limit).
\\[0.3cm]
Remark: In order that an observable qualify as a macro observable,
some properties have to be fulfilled
(cf. e.g. \cite{Kampen}).\vspace{0.3cm}

Typically a macro observable is the sum over few-body micro variables
(cf. \cite{Ludwig3} or \cite{Requ3}, see also the last section) like
e.g.
\begin{equation}A=c_f^{-1}\sum_{partitions}a(q_{i_1},\ldots,q_{i_n})\quad
  ,\quad f\gg n \end{equation} with the sum extending over all
partitions of $(1,\ldots,f)$ into $n$-element subsets and the constant
$c_f$ is of the order $f$. It can be shown that such observables
fulfill the above assumptions.

One can now represent an arbitrary state vector $\Psi(q)$ as a sum
over this new basis, i.e.
\begin{equation}\Psi(q)=\sum b_{Ji}\Phi_{Ji}(q)=\sum \Psi_J(q)     \end{equation}
with $\psi_J:=\sum_i b_{Ji}\Phi_{Ji}$. In a next step we will
introduce \tit{coarse} observables
$\overline{E},\overline{A},\overline{B},\dots$ with the property
\begin{equation}\overline{A}\,\Phi_{Ji}=A_J\,\Phi_{Ji}\quad ,\quad
  \overline{A}=\sum_{Ji}A_J\cdot P_{Ji}=\sum_J A_JP_J    \end{equation}
with $P_J=\sum_i P_{Ji}$. I.e., the $\Phi_{Ji}$ are now exact common eigenvectors of the
commuting set
$\{\overline{E},\overline{A},\overline{B},\dots\}$.\\[0.3cm]
Remark: Note that the existence of such observables is guaranteed by
the explicit construction via the above spectral representation.   
\vspace{0.3cm}

The expectation of e.g. $\overline{A}$ in the state $\Psi(q)$ is 
\begin{equation}\langle\Psi|\overline{A}|\Psi\rangle =\sum_J A_J\left(\sum_i|b_{Ji}|^2\right)=:\sum_JA_J\,w_J    \end{equation}
with $w_J$ the probability that the (macro) system is found in the
phase cell $J$. Furthermore $\overline{A}\Psi=\sum A_J\Psi_J$  
\begin{ob}The $w_J$ fix the macroscopic properties of the state
  $\Psi(q)$. \end{ob}

As to the technical details of the construction of such a set of macro
observables see the above cited papers. We give only one technical
property. 
\begin{ob}With $A=c_f^{-1}\sum_ka_k$, $B=c_{f'}^{-1}\sum_{k'}b_{k'}$, $a_k,b_{k'}$  microscopic few-body
  observables, we have
\begin{equation}[A,B]=c_f^{-1}\cdot (c'_f)^{-1}  \sum_{kk'}
  [a_k,b_{k'}]\approx 0 \quad\text{for}\quad f\gg 1   \end{equation}
\end{ob}
Proof: Note that by assumption most of the $[a_k,b_{k'}]\approx 0$,
that is, the set of terms, $[a_k,b_{k'}]$, being essentially different
from zero is of cardinality $O(f)$ and that $c_f=O(f)$. Hence
$c'_f\cdot c_f=O(f^2)$\\[0.3cm]
A fortiori, a macro observable (almost) comutes with all micro
observables in the large $f$-limit.
\begin{conclusion}Within the framework of true (many-body) quantum
  mechanics we found a subset of observables $E,A,B,\ldots$ which
  behave almost macroscopic, while the coarse observables
  $\{\overline{E},\overline{A},\overline{B},\dots\}$ exactly commute
  and have the common set of eigenvectors $\Phi_{Ji}$ which come in
  groups indexed by $J$. The macroscopic eigenvalues
  $E_J,A_J,B_J,\ldots$ are macroscopically discernible for $J\neq J'$
  with $\Phi_{Ji}$ having the macroscopic properties belonging to the
  group of micro states indexed by $J$.
\end{conclusion}

In the above approach we assumed (for convenience) that the spectrum
of the observables under discussion is discrete. In case we have an
observable with continuous spectrum the approach only needs a few
technical modifications. On the one hand, we can form observables with
discrete spectrum from observables with continuous spectrum by
appropriate coarse-graining. In a next step we can e.g. via rescaling
construct macro observables with (almost) continuous spectrum. As
example take certain position observables of some (macroscopic)
subsystems (cf. for example the last section).
\section{The Quantum Mechanical Measurement Process in the Light of
  the preceeding Analysis}
Our notion of macro systems and macro observables emerges as a
subtheory living on a second (coarse grained) level relative tothe
underlying quantum level. Furthermore, the algebra of macro
observables is a subalgebra of the full algebra of quantum
observables. I.e., in the space of microscopic observables we
(rigorously) construct a subspace of macro observables, $\mcal{A}_M$,
the members of which almost commute while the corresponding
coarse-grained observables, $\overline{\mcal{A}}_M$, exactly commute
by construction. The measurement devices or the pointers are assumed
to be essentially macroscopic, that is, pointer states or pointer
observables are assumed to belong to this class. This point of view is
in sharp contrast to the decoherence philosophy, where macroscopic
pointer states are assumed to be fixed via interaction with the
environment.
\begin{ob}[Superposition Principle]
  With $\Psi_1,\Psi_2$ many-body quantum states of a measurement
  instrument or of some macroscopic part of it (pointer), which are
  assumed to have unique macroscopic properties, i.e. belonging to
  single but different phase cells, that is
\begin{equation}\Psi_1(q)=\sum_i b_{Ji}^1\Phi_{Ji}(q)\quad ,\quad
\Psi_2(q)=\sum_i b_{J'i'}^2\Phi_{J'i'}(q)      \end{equation}
we have
\begin{equation}\Psi:= \Psi_1+\Psi_2= \sum_ib_{Ji}^1\Phi_{Ji}(q)+ \sum_{i'}b_{J'i'}^2\Phi_{J'i'}(q)\end{equation}
and (with $A_J$ the eigenvalues of some $\overline{A}_M$)
\begin{equation}\left(\Psi|\overline{A}_M|\Psi\right)=
  \left(\sum_i|b^1_{Ji}|^2\right) \cdot A_J+\left(\sum_{i'}|b^2_{J'i'}|^2\right) \cdot A_{J'}=\left(\Psi_1|\overline{A}_M|\Psi_1\right)+\left(\Psi_2|\overline{A}_M|\Psi_2\right)     \end{equation}
That is, within the realm of the smaller algebra $\overline{\mcal{A}}_M$ states
like $\Psi_l$ or $\Psi$ behave as mixtures and not as pure states.
\end{ob}

Note that in our approach the system is treated as a true quantum many-body
system in the microscopic regime and at the same time as a macro
system with respect to the smaller algebra $\mcal{A}_M$. This answers
(in our view) also some longstanding questions as to a possible
threshold where quantum properties go over (in a presumed
phase-transition-like manner) into macro properties. According to our
analysis there is no such threshold. It is rather the many-body
behavior as such which enables the selection of a subalgebra
$\mcal{A}_M$!
\begin{ob}[Schroedinger's Cat]
  The above result concerns superpositions of macro states being
  observably different, a catchword being\\ Schroedinger's Cat. In
  many discussions the wrong picture is invoked as if a superposition
  of dead and alive is something like a macroscopically blurred
  state. This impression is incorrect! What can be macroscopically
  observed is given by the class of macroscopic observables. But as we
  have shown, these observables annihilate the respective interference
  terms. Such interference terms could possibly only be observed in
  some super cosmos with the help of observables which connect
  macroscopically many degrees of freedom at a time (cf. the last
  section).
\end{ob}
Remark: We want to briefly comment upon a typical missconception
frequently ocurring in the literature. In the discussion of the
Schroedinger cat paradox the superposition of a dead and alive cat is
typically invoked. But this will actually never happen. On an
intermediate stage of the experiment the decay of e.g. a radioactive
atom and the fate of the cat is connected by the proper action of a
chain of various many-body devices setting ultimately free the
poison. That is, the macro state of the cat happens to be well
separated from the transition zone between micro and macro physics
with which our discussion is concerned.\vspace{0.3cm}

In a next step we want to show that the basis ambiguity problem
becomes obsolete in our context. 
\begin{ob}As all elements of $\mcal{A}_M$ quasi-commute or rigorously
  commute in $\overline{\mcal{A}}_M$, there do not exist the so-called
  complementary observables.
\end{ob}
This has the following effect. In e.g. the Stern-Gerlach experiment we
can of course formally repeat the analysis of Zurek and generalize
\begin{equation}1/\sqrt{2}\left((\uparrow)(\downarrow)-(\downarrow)(\uparrow)\right)=
  -1/\sqrt{2}\left(\psi_1'\psi_2'-\psi_2'\psi_1'\right) \end{equation}
(cf. section 2) to
\begin{equation}1/\sqrt{2}\left((\uparrow)\Phi_2-(\downarrow)\Phi_1\right)=     -1/\sqrt{2}\left(\psi_1'\Phi_1'-\psi_2'\Phi_2'\right)       \end{equation}
with $\Phi_1,\Phi_2$ some pointer states and
\begin{equation}\psi_1=1/\sqrt{2}\left((\uparrow)+(\downarrow)\right)\quad ,\quad \psi_2'=1/\sqrt{2}\left((\uparrow)-(\downarrow)\right)      \end{equation}
\begin{equation}\Phi_1'= 1/\sqrt{2}\left(\Phi_1-\Phi_2\right)\quad ,
  \quad \Phi_2'=1/\sqrt{2}\left(\Phi_1+\Phi_2\right)      \end{equation}
 I.e., the superposition principle is
taken for granted. However, there does not! exist a \tit{macro
  observable}, $\overline{B}_M$, so that the new states, $\Phi'_i$, are its
eigenstates. That is, the $\Phi_i'$ do not correspond to different
macroscopic pointer positions but rather are general many-body states
without distinct macroscopic properties. Macroscopically they rather represent
mixtures of macro states (cf. observation 4.1)

More precisely, we have
\begin{ob}With $\Phi_J$ eigenstates of the coarse macro observable
  $\overline{A}$, i.e. belonging to some phase cells $\mcal{C}_J$, 
\begin{equation}\overline{A}\circ\Phi_J=A_J\cdot\Phi_J    \end{equation}
there does not exist another coarse observable $\overline{B}$ with e.g.
\begin{equation}\overline{B}\circ\left(\Phi_1+\Phi_2\right)=B_3\cdot\left(\Phi_1+\Phi_2\right)    \end{equation}
that is, with $\left(\Phi_1+\Phi_2\right)$ another macroscopic state
with macroscopic properties. The state $\Phi_1+\Phi_2$ rather represents a mixture with respect to $\overline{\mcal{A}}_M$.
\end{ob}
Proof: 
\begin{equation}\overline{B}\circ\left(\Phi_1+\Phi_2\right)=B_3\cdot\left(\Phi_1+\Phi_2\right)    \end{equation}
implies (by assumption and definition) that $\Phi_1+\Phi_2$ is a macro
state. As all the macro observables commute, it is also an eigen state
of $\overline{A}$ with 
\begin{equation}\overline{A}\circ\left(\Phi_1+\Phi_2\right)=A_3\cdot\left(\Phi_1+\Phi_2\right)    \end{equation}
But we have 
\begin{equation}\overline{A}\circ\Phi_1=A_1\cdot\Phi_1\quad , \quad \overline{A}\circ\Phi_2=A_2\cdot\Phi_2   \end{equation}
with $A_1\neq A_2$. Hence
\begin{equation}\overline{A}\circ\left(\Phi_1+\Phi_2\right)\neq A_3\cdot\left(\Phi_1+\Phi_2\right)    \end{equation}
i.e., we get a contradiction, that is, $\Phi_1+\Phi_2$ can never be a
macro state with macroscopically definite properties.
\begin{conclusion}The basis ambiguity does not exist for
  $\overline{\mcal{A}}_M$. We can of course represent some many-body
  state with respect to another basis but the macroscopic properties
  remain the same! They are encoded in the a priori fixed
  decomposition 
\begin{equation}\Psi=\sum b_{Ji}\cdot\Phi_{Ji}\quad ,\quad \overline{A}\Phi_{Ji}=A_J \Phi_{Ji}    \end{equation}
and are the consequence of the complex many-body spectrum and the
reduced size of the algebra of macro observables. The entanglement
with the environment does not play a role in this analysis.
\end{conclusion}

In physical terms we can explain this result with the help of the
Stern-Gerlach experiment, following Bohr's dictum that the quantum
mechanical measurement of two complementary observables as
e.g. $\sigma_z$ and $\sigma_x$ need two different! and mutually
exclusive experimental setups. That is, in order to measure the
$z$-component one has to split the beam along the $z$-axis. This
implies that the magnets have to be oriented accordingly. The same
procedure with respect to the $x$-direction implies the respective
orientation of the magnets parallel to the $x$-axis. That is, we have
to apply a macroscopic rotation of the magnets (a many-body transformation).
\begin{ob}This rotation cannot be described by means of a
  superposition of states of the magnets being oriented in the
  $z$-direction as was the case in microscopic quantum mechanics of
  some spin variable. 
\end{ob}
\begin{conclusion}We can infer that the macroscopic pointer states are
  not determined via interaction (by decoherence) with the
  environment. They are obviously fixed a priori by the concrete
  experimental setup as described in the above example.
\end{conclusion}
\section{The Analysis of a Concrete Measurement Situation}
We now want to give a concrete example illuminating the approach,
described above. It was already essentially given in
\cite{Requ2},\cite{Requ3}. We assume that the pointer of our
measurement instrument is a macroscopic subsystem consisting of $N$
($N\gg 1$) quantum particles (e.g. a solid state system or an
avalanche in a Geiger-counter), being capable of performing
approximately a \tit{coherent motion}, depending on the micro state of
the quantum system to be measured.

I.e., in a concrete individual measurement event, the pointer as a
whole starts to move with a macroscopic momentum
\begin{equation}\left(\Phi_i(t)|\widehat{P}|\Phi_i(t)\right)\approx
  N\cdot <p>_i  \end{equation}
with $\widehat{P}=\sum_{i=1}^N \hat{p}_i$ the quantum mechanical total
momentum observable, $\Phi_i(t)$ a collective state of the pointer
(induced by the contact with the micro object) and 
\begin{equation} <p>_i:=
  \left(\Phi_i(t)|N^{-1}\cdot\widehat{P}|\Phi_i(t)\right) \end{equation}
the (approximately constant) \tit{mean-momentum} per (quantum-)
particle of the pointer. We assume that different measurement results
imply $<p>_i\neq <p>_j$ with the $<p>_i$ being in correspondence with
microscopic values $q_i$ of some quantum observable to be measured.

The \tit{center-of-mass} observable of the pointer
\begin{equation}\widehat{R}_{CM}:=\sum_i m_i\hat{r}_i/\sum_i m_i  \end{equation}
then behaves as (with $M:=\sum_i m_i$)
\begin{equation}\langle \widehat{R}_{CM}\rangle_i(t):=
  \left(\Phi_i(t)|\widehat{R}_{CM}|\Phi_i(t)\right)\approx const. +
  t\cdot <p>_i\cdot N/M   \end{equation}
\begin{ob}i) For $N\gg 1$ and $<p>_i\neq<p>_j$ the states
  $\Phi_i(t),\Phi_j(t)$ become (almost) orthogonal for macroscopic
  $t$.\\
ii) In our simple model the values $\{<p>_i\}$ label different phase
cells (or sectors) with (almost) sharp eigen values of the macro
observables $N^{-1}\cdot\sum_i\hat{p}_i$ or $\widehat{R}_{CM}$.\\
iii) An arbitrary microscopic state vector of our pointer system is a
superposition of the above sector states, i.e.
\begin{equation}\Psi=\sum_{Ji}b_{Ji}\Phi_{Ji}     \end{equation}
with $i$ labelling the different vectors belonging to the same phase
cell descibed by $<p>_J$.
\end{ob} 
\section{Interference among Macro States}
In this section we want to analyse the possibility of the observation
of interference effects among macroscopically distinct macro
states. We adressed this problem already in \cite{Requ2} and
\cite{Requ3}. We have shown in the preceeding analysis that this
cannot happen at a fixed time, $t$, in the regime of macro observables, $\mcal{A}_M$ or
$\overline{\mcal{A}}_M$. If one goes into the technical details one
observes that one (crucial) property, in order to qualify as a macro
observable, is the following
\begin{ob}With $N$ the number of microscopic constituents of a
  macroscopic (many-body) system ($N\gg 1$), we see that typical
  microscopic quantum mechanical observables are so-called
  \tit{few-body} observables. I.e.
\begin{equation}\hat{a}(x_{i_1},\ldots ,x_{i_n})       \end{equation}
denotes a microscopic $n$-particle observable, correlating $n\ll N$
microscopic constituents at a time. A typical many-body observable
which qualify as a macroscopic observable can then be written as
\begin{equation}\widehat{A}:=\sum_{Per}\hat{a}(x_{i_1},\ldots ,x_{i_n})      \end{equation}
where the sum extends over all possible clusters of $n$ micro objects
out of the $N$ constituents of the many-body system. Furthermore, a
prefactor of the order $N^{-1}$ frequently occurs in front of the sum.
\end{ob}

If we try to observe now possible off-diagonal elements of
$\widehat{A}$, that is, expectation values between different macro
states, we get approximately, making certain simplifying assumptions
\begin{ob}The degree of overlap between different macro states with
  respect to the macro observable $\widehat{A}$ is approximately
\begin{equation}|\left(\Phi_i|\widehat{A}|\Phi_j\right)|\approx
  N!/n!(N-n)!\cdot \tau^{(N-n)}      \end{equation}
with $\tau$ a small number ($\ll 1$) which denotes the individual
overlap of the wave function relative to the same microscopic
constituents in the different macro states, which do not! belong to
the cluster, coupled in a contribution coming from
e.g. $\hat{a}(x_{i_1},\ldots ,x_{i_n})$.
\end{ob}
\begin{conclusion}Interference between macroscopically different macro
  states could only be observed, if we were able to construct
  observables which do correlate $n\approx N\gg 1$ microscopic
  constituents at a time. The observables we are usually using in
  physics have however $n\ll N$. A situation where $n\approx N$ holds,
  is called by Ludwig in \cite{Ludwig3} a super-macro-cosmos.
\end{conclusion}  

We now come to an intersting point which was not treated in the preceding analysis, i.e., the possibility that the Hamiltonian time evolution generates interference effects, that cannot be observed at a fixed time $t$. This possibility was for example mentioned by Legett in \cite{Legett1} and \cite{Garg}. This possibility does not contradict our previous analysis which dealt with superpositions of different macro states at an arbitrary but fixed time. But, to say it in plain words, a superposition of e.g. two distinct macro states, $\Phi_1+\Phi_2$, which behaves at time $t=0$, according to our analysis, macroscopically as a mixture of $\Phi_1$ and $\Phi_2$ can, due to the Hamiltonian time evolution, evolve into another macro state, $\Phi(t)$, at some $t\neq 0$ which is not the mixture of the separate evolution of  $\Phi_1$ and $\Phi_2$.

To give a concrete example, we assume that $\Phi_1,\Phi_2$ are spatially separate at $t=0$ ($(\Phi_1,\Phi_2)=0$) so that no interference effects can be observed at time $t=0$. But at a later time they may display a marked overlap so that $(\Phi_1(t),\Phi_2(t))\neq 0$. This can easily be achieved for micro states (using e.g. a beam splitter, mirrors and/or a magnetic field) and there is in our view no a priori reason why the same cannot be accomplished for macroscopic (many-body) wave functions.
\begin{conclusion}It may happen that we have at time $t=0$ a superposition $\Phi_1+\Phi_2$ of e.g. two macroscopically distinct states which hence behaves like a mixture of $\Phi_1$ and $\Phi_2$  with respect to the algebra of macro observables. But at a later time $\Phi (t)=\Phi_1(t)+\Phi_2(t)$ is different from the mixture of $\Phi_1(t)$ and $\Phi_2(t)$.
\end{conclusion}

With the notations of \cite{Kampen} we can represent this more
quantitatively. Let the state at $t=0$ be
\begin{equation}\Psi(0)=\sum_{Ji}b_{Ji}\Phi_{Ji}     \end{equation}
The time evolution $U(t)$ leads to
\begin{equation}\Psi(t)=U(t)\circ \Psi(0)= \sum_{Ji}b_{Ji}U(t)\circ\Phi_{Ji}    \end{equation}
This equals
\begin{multline}\sum_{Ji}\sum_{J'i'}b_{Ji}(0)<\Phi_{J'i'},U(t)\Phi_{Ji}>\Phi_{J'i'}=
  \sum_{Ji}(\sum_{J'i'}b_{J'i'}(0)<\Phi_{Ji},U(t)\Phi_{J'i'}>)\Phi_{Ji}\\=\sum_{Ji}b_{Ji}(t)\Phi_{Ji}   \end{multline}
with 
\begin{equation}<\Phi_{Ji},U(t)\Phi_{J'i'}>=\sum_n
  <\Phi_{Ji},\psi_n>\cdot e^{-iE_nt/\hbar}\cdot
  <\psi_n,\Phi_{Ji}> \end{equation} and $\psi_n$ the eigenstates of
the (many-body) Hamiltonian. We hence have
\begin{multline}\sum_i|b_{Ji}(t)|^2= w_J(t)=\\\sum_{J'i',J''i''}(\sum_i<\Phi_{Ji},U(t)\Phi_{J'i'}>\overline{<\Phi_{Ji},U(t)\Phi_{J''i''}>})\cdot b_{J'i'}(0)\overline{b}_{J''i''}(0)    \end{multline}
\begin{ob}The $w_J(t)$ are in general not completely defined by the
  $w_J(0)$.
\end{ob}

In order to get more manageable formulas we will make
(cf. \cite{Kampen}) two \tit{dissorder assumptions}.
\begin{assumption}[Dissorder]
\begin{equation}w_J(t)\approx\sum_{J'i'}(\sum_i|<\Phi_{Ji},U(t)\Phi_{J'i'}>|^2)\cdot |b_{J'i'}(0)|^2 \end{equation}
and
\begin{equation}|b_{Ji}(0)|^2\approx \sum_i|b_{Ji}(0)|^2/D_J =w_j(0)/D_J\end{equation}
with $D_J$ the dimension of the subspace (phase cell) belonging to
$J$.
\end{assumption}  

With these (statistical) assumptions we get 
\begin{conclusion}Macroscopically we get
\begin{equation}w_J(t)=\sum_{J'}T_{JJ'}(t)w_{J'}(0)      \end{equation}
with
\begin{equation}T_{JJ'}(t)=D_J^{-1}\cdot\sum_{ii'}|<\Phi_{Ji},U(t)\Phi_{J'i'}>|^2     \end{equation}
That is, if the microscopic time evolution, induced by $H$, couples
micro states lying in different phase cells, $w_J(t)$ does not
depend only on $w_J(0)$. However, if it couples only micro states
coming from the same phase cell, $T_{JJ'}(t)$ is diagonal and we have
\begin{equation}w_J(t)=T_{JJ}(t)w_J(0)     \end{equation}
\end{conclusion}
\begin{koro}We see that in order that a measuring apparatus functions
as expected, we have to assume or, rather, to guarantee that within
the observation time the macroscopic time evolution is diagonal in the
above sense. One should note that this is the usual behavior of such
mechanical devices anyhow.
\end{koro}

\section{Conclusion}
We have shown that one can rigorously construct a subalgebra of
commuting macro observables within the set of quantum observables of a
generic many-body system. The common (almost) eigen values of this set
of macro observables are then the macroscopic properties of the
many-body system. Furthermore, for the subalgebra of macro observables
the basis ambiguity is lost (no complementarity!) and there is hence
no need for a (measurement-like) decoherence-by-environment mechanism
to fix the so-called pointer basis. The pointer basis is in our
approach already apriori fixed by the design of the measurement
instrument and by the spectral properties of the corresponding
microscopic many-body Hamiltonian together with the structure of the
set of macro observables.

\end{document}